\documentclass[aps,titlepage,12pt]{revtex4}
\usepackage{graphicx}
\usepackage{amsmath}
\usepackage{amssymb}
\usepackage{times}
\usepackage{color}
\usepackage{url}
\begin{document}

\title{Power of individuals -- Controlling centrality of temporal networks}
\author{Yujian Pan, Xiang Li$^{\dag}$}
\affiliation{Adaptive Networks and Control Laboratory, Department of Electronic Engineering, Fudan University, Shanghai 200433, China\\$\dag$Email: lix@fudan.edu.cn}

\begin{abstract}\noindent
\\
Temporal networks are such networks where nodes and interactions may appear and disappear at various time scales. With the evidence of ubiquity of temporal networks in our economy, nature and society, it's urgent and significant to focus on structural controllability of temporal networks, which nowadays is still an untouched topic. We develop graphic tools to study the structural controllability of temporal networks, identifying the intrinsic mechanism of the ability of individuals in controlling a dynamic and large-scale temporal network. Classifying temporal trees of a temporal network into different types, we give (both upper and lower) analytical bounds of the controlling centrality, which are verified by numerical simulations of both artificial and empirical temporal networks. We find that the scale-free distribution of node's controlling centrality is virtually independent of the time scale and types of datasets, meaning the inherent heterogeneity and robustness of the controlling centrality of temporal networks.
\end{abstract}

\maketitle

The recent outbreak of the A(H7N9) bird flu has caused much panic in China, and most of us still remember the financial crisis stretching from the USA to the world just a few years ago. These two impressive events are typical examples of complex networks in our economy, nature and society. Fortunately, considerable efforts have been dedicated to discovering the universal principles how structural properties of a complex network influence its functionalities~\cite{watts1998nature,barabasi1999science,albert2002rmp,newmani2006}. Not limited to understanding these statistical mechanics, another urgent aspect is to improve the capability to control such complex networks~\cite{wang2002physciaA,li2004ieee,li2006tac,yu2009automatica,rahmani2009siam,gutierrez2012sr}, and recent years have witnessed the blossoming studies on structural controllability of complex networks ~\cite{lombadi2007pre,liu2011nature,wang2012pre,nepusz2012np,yan2012prl,cowan2012plosone,posfai2013sr,liu2012plosone,delpini2013sr,sun2013prl,jia2013sr}. Classically, a linear time-invariant (LTI) dynamical system is controllable if, with a suitable choice of inputs, it can be driven from any initial state to any desired final state within the finite time~\cite{kalman1963,luenberger1979,slotine1991}. Structural controllability of a linear time-invariant system, initiated by Lin~\cite{lin1974tac} and further developed by other researchers~\cite{shields1976tac,hosoe1980tac,mayeda1981tac,poljak1989,poljak1990tac}, assumes free (non-zero) parameters of matrices $A^{'}$ and $B$ in
\begin{equation}
\dot{x}(t)=A^{'}x(t)+Bu(t)
\end{equation}
cannot be known exactly, and may attain some arbitrary but fixed values. A directed network, denoted as $G(A,B)$, associated with the above LTI system $(A^{'},B)$ is said to be structurally controllable, if $(A^{'},B)$ is controllable with the existence of matrices $\widetilde{A}$ and $\widetilde{B}$ structurally equivalent to $A^{'}$ and $B$, respectively. Noting that matrices $\widetilde{A}$ and $\widetilde{B}$ can be arbitrarily close to $A^{'}$ and $B$ when $(A^{'},B)$ is structurally controllable, and structural controllability is a general property in the sense that almost all weight combinations of a given network are controllable, except for some pathological cases with zero measure that occur when the parameters satisfy certain accidental constrains~\cite{liu2011nature,lin1974tac,shields1976tac}. In the existing literatures~\cite{lombadi2007pre,liu2011nature}, extensive efforts have been focused on the minimum number of input signals of such a network. Based on Lin's structural controllability theorem~\cite{lin1974tac}, Liu~\emph{et al.}~\cite{liu2011nature} stated that the minimizing problem can be efficiently solved by finding a maximum matching of a directed network, regarding a topologically static network as a linear time-invariant system. That is to say, a maximum subset of edges such that each node has at most one inbound and at most one outbound edge from the matching, and the number of nodes without inbound edges from the matching is the number of input signals required for maintaining structural controllability. With the minimum input theorem, many contributions to structural controllability of complex networks have been presented~\cite{wang2012pre,liu2012plosone,nepusz2012np,yan2012prl,cowan2012plosone,posfai2013sr,delpini2013sr,sun2013prl,jia2013sr}. Wang~\emph{et al.}~\cite{wang2012pre} proposed to optimize the structural controllability by adding links such that a network can be fully controlled by a single driving signal. Liu~\emph{et al.}~\cite{liu2012plosone} further introduced the control centrality to quantify the controllability of a single node. Nepusz~\emph{et al.}~\cite{nepusz2012np} evaluated the controllability properties on the edges of a network. Besides, controlling energy~\cite{yan2012prl}, effect of correlations on controllability~\cite{posfai2013sr}, evolution of controllability~\cite{delpini2013sr}, controllability transition~\cite{sun2013prl} and controlling capacity~\cite{jia2013sr}, have flourished very recently.

In our daily life, many networks fundamentally involve with time. The examples include the information flow through a distributed network and the spread of a disease in a population. Development of digital technologies and prevalence of electronic communication services provide a huge amount of data in large-scale networking social systems, including face-to-face conversations~\cite{isella2011jtb,takaguchi2012njp}, e-mail exchanges and phone calls~\cite{barabasi2005nature,gonzalez2008nature} and other types of interactions in various online behaviors~\cite{zhang2012epl,zhang2013chaos}. Such data are collectively described as temporal networks at specific time scales, where time-stamped events, rather than static ones, are edges between pairs of nodes (i.e. individuals)~\cite{holme2012pr}. More and more evidences indicate that the temporal features of a network significantly affect its topological properties and collective dynamic behaviors, such as distance and node centrality~\cite{kostakos2009physicaA,kim2012pre}, disease contagion and information diffusions~\cite{grindrod2011pre,perra2012prl}, characterizing temporal behaviors and components~\cite{tang2010pre,zhang2012,nicosia2012chaos} and scrutinizing the effects and characteristics within different time resolutions~\cite{ribeiro2012arxiv,krings2012epj,perra2012sr},
which are interdependent on the edge activations of temporal networks. However, to our best knowledge, a systematic study on structural controllability of temporal networks is still absent. In this paper, similar to the description of a static network by a LTI system~\cite{lin1974tac,liu2011nature}, a temporal network is associated with a linear time-variant (LTV) system as:
\begin{equation}
\dot{x}(t)=A^{'}(t)x(t)+B(t)u(t)
\end{equation}
where $A^{'}(t)\in\mathbb{R}^{N\times N}$ denotes the transpose of the adjacency matrix of a temporal network, i.e.,
$A^{'}(t)=(A(t))^{T}$, $x(t)=(x_{1}(t), x_{2}(t), \cdots, x_{N}(t))^{T}\in\mathbb{R}^{N}$ captures the time-dependent vector of the state variables of nodes,
$B(t)\in\mathbb{R}^{N\times M}$ is the so-called input matrix which identifies how external signals are fed into the nodes of the network, and
$u(t)=(u_{1}(t), u_{2}(t), \cdots, u_{M}(t))^{T}\in\mathbb{R}^{M}$ is the time-dependent input vector imposed by the outside controllers.
Meanwhile, by finding and classifying \emph{Temporal Trees} of a temporal network into different types with a combinational method of graph theory and matrix algebra, we introduce an index as the so-called controlling centrality to quantify the ability of a single node in controlling the whole temporal network. With analytical and experimental bounds, we point out the independence of the distribution of this centrality over different time scales. Besides, our method reserves as much temporal information as possible on structural controllability of temporal networks, which may shade new light on the study of structural controllability without wiping out information of the temporal dimension.\\

\noindent \textbf{\large Results} \\
\noindent A temporal network may include a sequence of graphs defined at discrete time points. Given a set of $N$ nodes, we denote the sequence of graphs as $\mathcal{G}=\{G^{t},t=1,2,\cdots,T\}$, where $T$ is the sequence length, and $G^{t}$ is a static graph sampled at time point $t$. The adjacency matrix of a temporal network, $\mathcal{G}$, can be denoted by a $N\times N$ time-dependent adjacency matrix $A(t)$, $t=1,2,\cdots,T$, where $a_{ij}(t)$ are the elements of the adjacency matrix of the $t^{th}$ graph, $G^{t}$.
\begin{table}
\begin{tabular}{|c|c|c|c|}
  \hline
  Node Pair(Contact) & Active Time Points & Node Pair(Contact) & Active Time Points \\
  \hline
  (A, B) & [1,2,3,4] & (B, C) & [4,6] \\
  (C, D) & [2,3] & (D, E) & [3,4,5,6] \\
  (E, F) & [1,3] & (B, F) & [5,6] \\
  (C, F) & [4,5,6] & ~~~~~~ & ~~~~~~\\
  \hline
\end{tabular}
\caption{The temporal network in Fig. 1 with the node pairs and active contacts}
\end{table}
For example, a temporal network, $\mathcal{G}$, with the set of contacts in Table I can be sampled as a sequence of graphs at time points $t=1,2,\cdots,6$,
denoted as $\mathcal{G}=\{G^{1}, G^{2}, G^{3}, G^{4}, G^{5}, G^{6}\}$ and shown in Fig. 1.
We illustrate the propagation process taking place on the temporal network as shown in Fig. 2. Actually, a message can only arrive at nodes \emph{B, C and F }(dotted nodes in Fig. 2) if its source is located on node \emph{A}, though each node can receive the same message if the source is located on node \emph{D}. This asymmetry (node \emph{D} reaches node \emph{A}, while not vice versa) mainly due to the direction of time evolution, highlights a fundamental gap between static and temporal networks.\\

\noindent \textbf{Structurally Controlling Centrality of Temporal Networks}
Generally, non-zero entries of a matrix $A$ are free, and $A$ is structured if the free entries are (algebraically) independent. Two matrices $A$ and $\widetilde{A}$ are same structured if their zero entries coincide. Matrices $A,B,C,\cdots$ are independent if all free entries of these matrices are (algebraically) independent. In particular, any independent matrix must be structured, and any two entries of two matrices must be distinct~\cite{lin1974tac,poljak1990tac}. A temporal network is said to be structurally controllable at time point $t_{0}$ if its associated LTV system described by Eq.(2), with a suitable choice of inputs $u(t)$, can be driven from any initial state to any desired final state within the finite time interval $[t_{0}, t_{1}]$ , where the initial and finial states are designated at time point $t_{0}$ and $t_{f}$ $(t_{0}<t_{f}\leq t_{1})$, respectively.

For simplicity, we focus on the case of a single controller and reduce the input matrix $B(t)$ in Eq. (2) to the input vector $b^{(o)}$ with only a single non-zero element, and rewrite Eq. (2) as
\begin{equation}
\dot{x}(t)=A^{'}(t)x(t)+b^{(o)}u(t)
\end{equation}

With non-periodic sampling of Eq. (3), we get its discrete version with the recursive relationship for any two neighboring state spaces of a temporal network
\begin{equation}
x(k+1)=G_{k+1}x(k)+H_{k+1}u(k), (k=0, 1, \cdots, T-1)
\end{equation}
Define $S_{M(o)}$ the structurally controlling centrality of node $o$ in a temporal network:
\begin{equation}
S_{M(o)}=rank(W_{c})=rank([G_{T}\cdots G_{2}H_{1}, \cdots, G_{T}H_{T-1}, H_{T}])
\end{equation}
where $G_{k+1}=I+T_{k+1}A_{k+1}^{'}$, $H_{k+1}=T_{k+1}b^{(o)}$, $A_{k+1}^{'}$ is the transpose of the adjacency matrix of the $(k+1)$th graph,
$I$ and $T_{k+1}=t_{k+1}-t_{k}$ are the identity matrix and the sampling interval, respectively. $S_{M(o)}$ is a measure of node $o$'s ability to
structurally control the network, i.e. the maximum dimension of controllable subspace (see Methods), and in this paper, $G_{k+1}$ and $H_{k+1}$ are structured matrices of size $N\times N$ and $N\times 1$, respectively.\\

\noindent \textbf{Graph Characteristics}
Given a temporal network $\mathcal{G}(V_{\mathcal{G}},E_{\mathcal{G}})$, where $V_{\mathcal{G}}$ and $E_{\mathcal{G}}$ are the collection of nodes and interactions, respectively, we associate $\mathcal{G}$ with another acyclic digraph $N(\mathcal{G},T)$. The vertex set of $N(\mathcal{G},T)$ consists of $T+1$ copies, i.e., $i_{1},i_{2},\cdots,$ and $i_{T+1}$, of each vertex $i\in V_{\mathcal{G}}$, and $T+1$ copies, i.e., $I^{o}_{0},I^{o}_{1},\cdots,$ and $I^{o}_{T}$, of the single controller $I^{o}$, denoted as the red ones in Fig. 3 (b). The edge set of $N(\mathcal{G},T)$ consists of three types of edges: (i) the edges connecting node $i$ at neighboring time points, i.e., $i_{t}\rightarrow i_{t+1},t=1,2,\cdots,T$, for each node $i\in V_{\mathcal{G}}$, (ii) the edges $i_{t}\rightarrow j_{t+1}$, where $i\rightarrow j\in E_{G^{t}},t=1,2,\cdots,T$ and (iii) the edges connecting the controller $I^{o}$, i.e., $I^{o}_{t}\rightarrow o_{t+1},t=0,1,\cdots,T$, where $o\in V_{\mathcal{G}}$ denotes the directly controlled node. These aforementioned three types of edges are denoted as the red dotted ones, the blue ones and the black ones in Fig. 3 (b), respectively.Such interpretation of a temporal network is called the Time-Ordered Graph (TOG) model in ~\cite{kim2012pre}, which transforms a temporal network into a larger but more easily analyzable static version. For example, we translate the temporal network of Fig. 3 (a) to the corresponding time-ordered graph as shown in Fig. 3 (b). With the TOG model, we first give the definition of input reachability in a temporal network.

\noindent \textbf{Definition 1:} Consider subset $S_{1}=\{I^{o}_{0},I^{o}_{1},\cdots,I^{o}_{T}\}$ and node $i_{T+1}\in S_{2}= \{1_{T+1},2_{T+1},\cdots,|V_{\mathcal{G}}|_{T+1}\}$ of $N(\mathcal{G},T)$, which correspond to node $I^{o}$ and node $i\in V_{\mathcal{G}}$ of $\mathcal{G}$, respectively. If in $N(\mathcal{G},T)$ there exists a path to $i_{T+1}$, whose tail $I^{o}_{t}\in S_{1}$, then node $i$ of $\mathcal{G}$ is reachable from node $I^{o}$ at time $t$, and the set of such reachable nodes in $V_{\mathcal{G}}$ is the reachable subset of the input signal $I^{o}$ of $\mathcal{G}$.

\noindent \textbf{Proposition 1:} The reachability of the input signal of $\mathcal{G}$ is equivalent to the reachability of subset $S_{1}$, i.e. the $t^{th}$ row of the $t^{th}$ power of adjacency matrix of $N(\mathcal{G},T)$, and the controlled rows of dynamic communicability matrices of $\mathcal{G}$ starting at different time points $t$, denoted as $\{Q_{t}\}_{o,\forall}$, where $t=1,2,\cdots,T$.

\noindent \emph{Proof:} Denote partitioned matrix $A_{N(\mathcal{G},T)}$ (size $(|V_{\mathcal{G}}|+1)\times (|V_{\mathcal{G}}|+1)$) as the adjacency matrix of $N(\mathcal{G},T)$, and for each block $B_{(i,j)}$ (size $(T+1)\times (T+1)$) of matrix $A_{N(\mathcal{G},T)}$, if there's a directed edge $i_{t}\rightarrow j_{t+1}$ in $N(\mathcal{G},T)$, where $t=1,2,\cdots,T$, then we have $\{B_{(i,j)}\}_{t,t+1}\neq0$ and $\{A_{N(\mathcal{G},T)}\}_{i(T+1)+t,j(T+1)+t+1}\neq0$. Recall the dynamic communicability matrix ~\cite{grindrod2011pre} to quantify how effectively a node can broadcast and receive messages in a temporal network, defined as :
\begin{equation}
Q:=(I+aA_{1})(I+aA_{2})\cdots(I+aA_{T})
\end{equation}
Here, matrix $A_{t}$ is the adjacency matrix of the $t^{th}$ graph, and $0<a<1/\rho$ ($\rho$ denotes the maximum spectral radius of matrices). Similarly, we define the communicability matrix starting at different time points to quantify the reachability of the controller, written as:
\begin{equation}
Q_{t}=(I^{*}+a_{t}A_{t}^{*})(I^{*}+a_{t+1}A_{t+1}^{*})\cdots(I^{*}+a_{T}A_{T}^{*})
\end{equation}
where $A_{t}^{*}=\begin{pmatrix}0 & (b^{(o)})^{'}\\
\mathbf{0}_{N\times 1} & A_{t}\end{pmatrix}$ is the adjacency matrix of the $t^{th}$ graph with a single controller $I^{o}$ located on node $o$, and
$I^{*}=\begin{array}{ll}
\begin{pmatrix}
0 & \mathbf{0}_{1\times N}\\ \mathbf{0}_{N\times 1} & I_{N\times N}
\end{pmatrix}
\end{array}$. Note that a non-zero element $(i,j)$ of a product of matrices, such as $(A)^{k}$, is the reachability from node $i$ to node $j$ if $\{(A)_{k}\}_{i,j}\neq0$, and the length of paths in graph $N(\mathcal{G},T)$ is never more than $T+1$. Therefore, the reachability of node $I^{o}_{t}$ in $S_{1}=\{I^{o}_{0},I^{o}_{1},\cdots,I^{o}_{T}\}$ is the $(t+1)^{th}$ row of $(T+1-t)^{th}$ power of $A_{N(\mathcal{G},T)}$, i.e., $\{(A_{N(\mathcal{G},T)})^{T+1-t}\}_{t+1,\forall}$, where $t=0,1,\cdots,T$. For each column of matrix $W_{c}$, we have $G_{T}\cdots G_{2}H_{1}=[(G_{T}\cdots G_{2})^{'}]^{'}H_{1}=[G_{2}^{'}\cdots G_{T}^{'}]^{'}H_{1}=[(I+A_{2})\cdots (I+A_{T})]^{'}H_{1}$, and with the definition of matrix $Q_{t}$, we know that $\{Q_{t}\}_{i,j}$ describes the reachability from node $i$ to node $j$. Therefore, the rechability of controller $I^{o}$ at time $t$ is equivalent to the controlled row, i.e. the $o^{th}$ row, denoted as $\{Q_{t}\}_{o,\forall}$, of matrix $Q_{t}$. $\blacksquare$

With \emph{Proposition 1}, we rewrite matrix $W_{c}$ in the form of reachability as:
\begin{equation}
W^{*}=[(\{Q_{1}\}_{o,\forall})^{'}, (\{Q_{2}\}_{o,\forall})^{'}, \cdots, (\{Q_{T}\}_{o,\forall})^{'}]=
\begin{pmatrix}
\mathbf{0}_{1\times T}\\
W_{c}
\end{pmatrix}
\end{equation}
where $\{Q_{t}\}_{1,\forall}$ denotes the reachability of the controller at time point $t$, and we have $rank(W_{c}^{*})=rank(W_{c})$. As shown in Fig. 3, we easily get $\{Q_{1}\}_{1,\forall}=[0,~1+a_{21}c_{31}b_{42},~a_{21},~a_{21}c_{31}+b_{41},~a_{21}d_{41}]$, $\{Q_{2}\}_{1,\forall}=[0,~1,~0,~b_{41},~0]$, $\{Q_{3}\}_{1,\forall}=[0,~1,~0,~b_{41},~0]$, and $\{Q_{4}\}_{1,\forall}=[0,~1,~0,~0,~0]$. According to \emph{Proposition 1}, $W^{*}=[(\{Q_{1}\}_{1,\forall})^{'},~(\{Q_{2}\}_{1,\forall})^{'},~(\{Q_{3}\}_{1,\forall})^{'},~(\{Q_{4}\}_{1,\forall})^{'}]=\begin{pmatrix}
0 & 0 & 0 & 0\\ 1+a_{21}c_{31}b_{42} & 1 & 1 & 1\\ a_{21} & 0 & 0 & 0\\ a_{21}c_{31}+b_{41} & b_{41} & b_{41}& 0\\ a_{21}d_{41} & 0 & 0 & 0
\end{pmatrix}$.

\noindent \textbf{Definition 2:} A temporal tree, denoted as $TT_{t}$, of a temporal network $\mathcal{G}(V_{\mathcal{G}},E_{\mathcal{G}})$ is a Breadth-First Search (BFS) spanning tree, denoted as $ST_{t}$, of its corresponding static network $N(\mathcal{G},T)$ (TOG model) rooted at node $I^{o}_{t}\in S_{1}=\{I^{o}_{0},I^{o}_{1},\cdots,I^{o}_{T}\}$.\\
\noindent \emph{Remark:} The Breadth-First Search (BFS) is a classical strategy for searching nodes in graph theory, and a BFS spanning tree contains all the nodes and edges when the BFS strategy is applied at some node. A distinctive property of $N(\mathcal{G},T)$ is that there's no cycles in it, and each path's length is no more than $T+1$, so it's easy to apply the BFS strategy to find trees rooted at some designated nodes in $N(\mathcal{G},T)$. Obviously, the one-one mapping between a temporal tree of a temporal network and a BFS spanning tree of the TOG is guaranteed by the one-one mapping between $\mathcal{G}(V_{\mathcal{G}},E_{\mathcal{G}})$ and $N(\mathcal{G},T)$. For the temporal network in Fig. 3 (a), each of the three temporal trees, as shown in Fig. 3 (c), of this temporal network exists a unique corresponding BFS spanning tree, as shown in Fig. 3 (d).

\noindent \textbf{Proposition 2:} Denote $R_{TT_{1}},R_{TT_{2}},\cdots,$ and $R_{TT_{T}}\in\mathbb{R}^{(V_{\mathcal{G}}+1)\times 1}$ as the reachability vector of each temporal tree from the controller $I^{o}$, and matrix $W^{R}=[R_{TT_{1}}~R_{TT_{2}}~\cdots~R_{TT_{T}}]\in\mathbb{R}^{(V_{\mathcal{G}}+1)\times T}$, we have $rank(W^{R})=rank(W^{*})$.

\noindent \emph{Proof:} With \emph{Proposition 1} and \emph{Definition 2}, we know there's a temporal tree $TT_{t}$ of each $ST_{t}$ in TOG, and each $ST_{t}$ is a leading tree when compared with $ST_{t+1}$ (refer to the definition of BFS spanning tree with the TOG model). Therefore, each temporal tree $TT_{t}$ is a leading tree when compared with $TT_{t+1}$. Two strategies are adopted to yield a leading temporal tree: i) Adding new nodes into $TT_{t}$, i.e., we have $|V_{TT_{t}}|>|V_{TT_{t+1}}|$, ii) Adding new paths to the existing nodes, i.e., we have $|V_{TT_{t}}|=|V_{TT_{t+1}}|$. In the case of strategy i), if there's only one temporal tree, we obviously have $rank(W^{R})=rank(R_{TT})=rank(W^{*})=1$; if the number of temporal trees is $n$, and $rank(W^{*})=rank(W^{R})=n$, then when the number of temporal trees is $n+1$, we have $rank(W^{R})=rank\begin{pmatrix}
\mathbf{0}_{1\times n} & 0\\W^{*}_{n\times n} & ($\ddag$)_{n\times 1}\\\mathbf{0}_{(|V_{\mathcal{G}}|-n)\times n} & ($\ddag$)_{(|V_{\mathcal{G}}|-n)\times 1}
\end{pmatrix}=rank(W^{*})=n+1$, where ($\ddag$) denotes a nonzero vector. In the case of strategy ii), each new interaction in leading tree $TT_{t}$, which isn't included in temporal tree $TT_{t+1}$, contributes to new paths to the existing nodes. By some linear superposition of columns of matrix $W^{*}$ and $W^{R}$, we find there's no impact on the maximum rank of matrix $W^{*}$ if we cut down and drop those "old" interactions, which means we only need to take the leading temporal tree, i.e. $TT_{t}$, into consideration. Therefore, we have $rank(W^{R})=rank(P_{1}W^{R}T_{1})=rank(W^{*})=rank(P_{2}W^{*}T_{2})$, where $P_{1}$, $P_{2}$, $T_{1}$ and $T_{2}$ are properly defined linear transformation matrices. $\blacksquare$

For example, according to \emph{Definition 2}, the reachability of temporal tree $TT_{1}$ of Fig. 3 (c) is $R_{TT_{1}}=[0,~1,~a_{21},~a_{21}c_{31},~a_{21}d_{41}]^{'}$. Similarly, we have $R_{TT_{2}}=[0,~1,~0,~c_{41},~0]^{'}$, $R_{TT_{3}}=[0,~1,~0,~c_{41},~0]^{'}$ and $R_{TT_{4}}=[0,~1,~0,~0,~0]^{'}$ for temporal trees $TT_{2}$, $TT_{3}$ and $TT_{4}$, respectively. Therefore, we easily reach $P_{1}W^{R}T_{1}=P_{1}[R_{TT_{1}},~R_{TT_{2}},~R_{TT_{3}},~R_{TT_{4}}]T_{1}=
P_{1}\begin{pmatrix}
0 & 0 & 0 & 0\\ 1 & 1 & 1 & 1\\ a_{21} & 0 & 0 & 0\\ a_{21}c_{31} & b_{41} & b_{41} & 0\\ a_{21}d_{41} & 0 & 0 & 0
\end{pmatrix}T_{1}=\begin{pmatrix}
0 & 0 & 0 & 0\\ 0 & 0 & 0 & 1\\ a_{21} & 0 & 0 & 0\\ a_{21}c_{31} & 0 & b_{41} & 0\\ a_{21}d_{41} & 0 & 0 & 0
\end{pmatrix}$, and $P_{2}W^{*}T_{2}=P_{2}\begin{pmatrix}
0 & 0 & 0 & 0\\ 1+a_{21}c_{31}b_{42} & 1 & 1 & 1\\ a_{21} & 0 & 0 & 0\\ a_{21}c_{31}+b_{41} & b_{41} & b_{41}& 0\\ a_{21}d_{41} & 0 & 0 & 0
\end{pmatrix}T_{2}=\begin{pmatrix}
0 & 0 & 0 & 0\\ 0 & 0 & 0 & 1\\ a_{21} & 0 & 0 & 0\\ a_{21}c_{31} & 0 & b_{41} & 0\\ a_{21}d_{41} & 0 & 0 & 0
\end{pmatrix}$. Obviously, $rank(W^{R})=rank(P_{1}W^{R}T_{1})=rank(W^{*})=rank(P_{1}W^{*}T_{1})$, which is consistent with \emph{Proposition 2}.

\noindent \textbf{Definition 3:} Temporal trees $TT_{1},TT_{2},\cdots$ are homogeneously structured if their corresponding adjacency matrices, denoted as $A_{TT_{1}},A_{TT_{2}},\cdots$, are same structured. Otherwise, they are heterogeneously structured.

We rewrite matrix $W^{R}$ as:
\begin{equation}
W^{R}=\begin{pmatrix}
W^{D} & W^{S}
\end{pmatrix}=
\begin{pmatrix}\mathbf{0}_{1\times T^{D}} & \mathbf{0}_{1\times T^{S}}\\W_{c}^{D} & W_{c}^{S}\end{pmatrix}
\end{equation}
In Eq. (9), matrix $W^{D}=\begin{pmatrix}
\mathbf{0}_{T^{D}\times 1}~~(W_{c}^{D})^{'}
\end{pmatrix}^{'}$ of size $(|V_{\mathcal{G}}|+1)\times T^{D}$ denotes the part of heterogeneously structured trees, and matrix $W^{S}=\begin{pmatrix}
\mathbf{0}_{T^{S}\times 1}~~(W_{c}^{S})^{'}
\end{pmatrix}^{'}$ of size $(|V_{\mathcal{G}}|+1)\times T^{S}$ denotes the part of homogeneously structured trees, respectively. Obviously, $T^{D}+T^{S}=T$.

\noindent \textbf{I. Heterogeneously Structured Trees:}

\noindent \textbf{Definition 4:} If heterogeneous trees $TT_{D1},TT_{D2},\cdots$ consist of same nodes, i.e., $V_{TT_{D1}}=V_{TT_{D2}}=\cdots$, then they are called heterogeneous trees with same nodes. Otherwise they are heterogeneous trees with different nodes.

To determine the rank of matrix $W^{D}$, we rewrite it as:
\begin{equation}
W^{D}=\begin{pmatrix}
\mathbf{0}_{1\times T_{S_{1}}^{D}} & \mathbf{0}_{1\times T_{S_{2}}^{D}} & \cdots & \mathbf{0}_{1\times T_{S_{k}}^{D}} & \vline & \mathbf{0}_{1\times T_{S_{k+1}}^{D}} \\
W^{D}_{S_{1}} & W^{D}_{S_{2}} & \cdots & W^{D}_{S_{k}} & \vline & W^{D}_{S_{k+1}}\\
\end{pmatrix}
\end{equation}
In Eq. (10), each $W^{D*}_{S_{l}}=\begin{pmatrix}
\mathbf{0}_{T_{S_{l}}^{D}\times 1}~~(W^{D}_{S_{l}})^{'}
\end{pmatrix}^{'}$, $l=1,2,\cdots,k$, of size $(|V_{\mathcal{G}}|+1)\times T_{S_{l}}^{D}$ denotes a collection of heterogeneous trees with same nodes ($V_{W^{D*}_{S_{l}}}\neq V_{W^{D*}_{S_{l^{'}}}}$ for $\forall l\neq l^{'}$), and $W^{D*}_{S_{k+1}}=\begin{pmatrix}
\mathbf{0}_{T_{S_{k+1}}^{D}\times 1}~~(W^{D}_{S_{k+1}})^{'}
\end{pmatrix}^{'}=[R_{TT_{1}}^{D}, R_{TT_{2}}^{D},\cdots, R_{TT_{T_{S_{k+1}}^{D}}}^{D}]$ of size $(|V_{\mathcal{G}}|+1)\times T_{S_{k+1}}^{D}$ denotes heterogeneous trees with different nodes. $\sum_{l=1}^{k+1} T_{S_{l}}^{D}=T^{D}$.

\noindent \textbf{Case 1:} Heterogeneous trees with same nodes.

\noindent \textbf{Proposition 3:} Given matrix $W^{D*}_{S_{l}}$ as a collection of heterogeneous trees with same nodes, we have $rank(W^{D*}_{S_{l}})=min(|V_{W^{D*}_{S_{l}}}|,T_{S_{l}}^{D})$, $l=1,2,\cdots,k$, where $|V_{W^{D*}_{S_{l}}}|$ denotes the number of nodes in matrix $W^{D*}_{S_{l}}$.

\noindent \emph{Proof:} According to the definition of heterogeneous trees with same nodes, these trees always have the same reachability with different paths to reach the same node, which means for each heterogeneously structured temporal tree with same nodes, there exists at least one independent parameter (interaction). When $T_{S_{l}}^{D}=1$, $rank(W^{D*[1]}_{S_{l}})=1$. When $T_{S_{l}}^{D}=n\leq |V_{W^{D*}_{S_{l}}}|$, we get a triangular matrix with its diagonal elements non-zeros by some linear transformations. Therefore, we have $rank(W^{D*}_{S_{l}})=T_{S_{l}}^{D}=n$. Similarly, when $|V_{W^{D*}_{S_{l}}}|\leq T_{S_{l}}^{D}=n$, we get $rank(W^{D*}_{S_{l}})=|V_{W^{D*}_{S_{l}}}|$. In short, we reach $rank(W^{D*}_{S_{l}})=min(|V_{W^{D*}_{S_{l}}}|,T_{S_{l}}^{D})$. $\blacksquare$

\noindent \textbf{Case 2:} Heterogeneous trees with different nodes.

\noindent \textbf{Proposition 4:} Given matrix $W^{D*}_{S_{k+1}}$ as heterogeneous trees with different nodes, we have $rank(W^{D*}_{S_{k+1}})=T_{S_{k+1}}^{D}$.

\noindent \emph{Proof:} When $T_{S_{k+1}}^{D}=1$, we easily have $rank(W^{D*[1]}_{S_{k+1}})=1$. If $T_{S_{k+1}}^{D}=n$, and $rank(W^{D*[n]}_{S_{k+1}})=n$, then when $T_{S_{k+1}}^{D}=n+1$, it's equivalent to add a tree with different nodes into matrix $W^{D*[n]}_{S_{k+1}}$ to get matrix $W^{D*[n+1]}_{S_{k+1}}$. Therefore, there always exists at least one new nonzero entry with its column index $n+1$ and row index $r (n<r\leq |V_{\mathcal{G}}|+1)$ in matrix $W^{D*[n+1]}_{S_{k+1}}$, and
 $rank(W^{D*[n+1]}_{S_{k+1}})=rank([W^{D*[n]}_{S_{k+1}},R_{TT}^{D}])=
rank\begin{pmatrix}
\mathbf{0}_{1\times n} & 0\\W^{*}_{n\times n} & ($\ddag$)_{n\times 1}\\\mathbf{0}_{(|V_{\mathcal{G}}|-n)\times n} & ($\ddag$)_{(|V_{\mathcal{G}}|-n)\times 1}
\end{pmatrix}=n+1$, where ($\ddag$) denotes a nonzero vector. That means for any $T_{S_{k+1}}$, we have $rank(W^{D*}_{S_{k+1}})=T_{S_{k+1}}^{D}$. Note that if we cannot find such a nonzero entry, we claim that this new tree must have a collection of nodes coincident to some other tree, which is not allowed in this case. $\blacksquare$

\noindent \textbf{Theorem 1:} Given matrices $W^{D*}_{S_{l}}$, $l=1,2,\cdots,k+1$, as the heterogeneously structured trees and $S^{D}_{M(o)}$ as the maximum-structurally controllable subspace of heterogeneously structured trees, we have
\begin{equation}
max_{l=1}^{k+1}\{rank(W^{D*}_{S_{l}})\}\leq S^{D}_{M(o)}=rank(W^{D})\leq \sum_{l=1}^{k+1} rank(W^{D*}_{S_{l}})
\end{equation}

\noindent \emph{Proof:} Firstly, we prove the left part of inequality (11), i.e. $rank(W^{D})\geq max_{l=1}^{k+1}\{rank(W^{D*}_{S_{l}})\}$. Compared with the trees, denote as $TT_{1}, TT_{2},\cdots, TT_{T_{S_{k+1}}^{D}}$, in matrix $W^{D*}_{S_{k+1}}$ ($W^{D*}_{S_{k+1}}=[R_{TT_{1}}^{D}, R_{TT_{2}}^{D},\cdots, R_{TT_{T_{S_{k+1}}^{D}}}^{D}]$), those trees in matrices $W^{D*}_{S_{l}}, l=1,2,\cdots,k,$ have different nodes, i.e., $V_{W^{D*}_{S_{l}}}\neq V_{TT_{1}}\neq V_{TT_{2}}\neq \cdots\neq V_{TT_{T_{S_{k+1}}^{D}}}$, and $V_{W^{D*}_{S_{l}}}\neq V_{W^{D*}_{S_{l^{'}}}}$ for $\forall l\neq l^{'}$. Therefore, $rank(W^{D})=max_{l=1}^{k+1}\{rank(W^{D*}_{S_{l}})\}$ when there exists a matrix consists of all nodes, and it has the maximum rank. For the right part, i.e. $rank(W^{D})\leq \sum_{l=1}^{k+1} rank(W^{D*}_{S_{l}})$, we reach the equality when matrix $W^{D}$ is written as: $W^{D}=\begin{pmatrix}
\mathbf{0}_{1\times T_{S_{1}}^{D}} & \mathbf{0}_{1\times T_{S_{2}}^{D}} & \cdots & \mathbf{0}_{1\times T_{S_{k+1}}^{D}}\\
1 & 1 & \cdots & 1\\
W^{*1} & \mathbf{0}_{r_{1}\times T_{S_{2}}^{D}} & \cdots & \mathbf{0}_{r_{1}\times T_{S_{k+1}}^{D}}\\
\mathbf{0}_{r_{2}\times T_{S_{1}}^{D}} & W^{*2} & \cdots & \mathbf{0}_{r_{2}\times T_{S_{k+1}}^{D}}\\
\vdots & \vdots & \ddots & \vdots\\
\mathbf{0}_{r_{k+1}\times T_{S_{1}}^{D}} & \mathbf{0}_{r_{k+1}\times T_{S_{k}}^{D}} &  \cdots & W^{*k+1}
\end{pmatrix}$, where row vector $\mathbf{1}_{1\times T^{D}}$, i.e, the $2^{th}$ row of matrix $W^{D}$, denotes node $o$, and matrices $W^{*1}, W^{*2},\cdots,W^{*k+1}$ denote the other part of these trees. This means there's no intersection of nodes between any two matrices of $W^{D*}_{S_{l}}, l=1,2,\cdots,k+1,$ except node $o$, i.e., $\mid V_{W^{D*}_{S_{l}}}\cap V_{W^{D*}_{S_{l^{'}}}}\mid=1$ for $\forall l\neq l^{'}$ .
In this case, each matrix $W^{D*}_{S_{l}}$ contributes $d=rank(W^{D*}_{S_{l}})= T_{S_{l}}^{D}$ to $rank(W^{D})$. Therefore, $rank(W^{D})= \sum_{l=1}^{k+1} rank(W^{D*}_{S_{l}})$. $\blacksquare$

\noindent \textbf{II. Homogeneously Structured Trees:}

\noindent \textbf{Definition 5:} Consider homogeneously structured trees $TT_{S1},TT_{S2},\cdots$. If their corresponding adjacency matrices $A_{TT_{S1}},A_{TT_{S2}},\cdots$ are independent, then they are called independent trees. Otherwise they are interdependent trees.

We rewrite matrix $W^{S}$ as:
\begin{equation}
W^{S}=\begin{pmatrix}
W^{S}_{1}~~W^{S}_{2}~~\cdots~~W^{S}_{q}
\end{pmatrix}
\end{equation}
and each $W^{S}_{m},(m=1,2,\cdots,q),$ denote a collection of homogeneously structured trees ($V_{W^{S}_{m}}\neq V_{W^{S}_{m^{'}}}$ for $\forall m\neq m^{'}$), which is written as:
\begin{equation}
W^{S}_{m}=\begin{pmatrix}
\mathbf{0}_{1\times T_{S_{m},1}^{S}} & \cdots & \mathbf{0}_{1\times T_{S_{m},p(m)}^{S}} & \vline & \mathbf{0}_{1\times T_{S_{m},p(m)+1}^{S}}\\
W_{S_{m},1}^{S} & \cdots & W_{S_{m},p(m)}^{S} & \vline & W_{S_{m},p(m)+1}^{S}\\
\end{pmatrix}
\end{equation}
In Eq. (13), each $W_{S_{m},w}^{S*}=\begin{pmatrix}
\mathbf{0}_{T_{S_{m},w}^{S}\times 1}~~(W_{S_{m},w}^{S})^{'}
\end{pmatrix}^{'}$, $w=1,2,\cdots,p(m)$, of size $(|V_{\mathcal{G}}|+1)\times T_{S_{m},w}^{S}$ denotes a collection of interdependent trees with same interactions ($I_{m,w}\neq I_{m,w^{'}}$ for $\forall w\neq w^{'}$, where $I_{m,w}$ denotes the collection of same interactions in matrix $W_{S_{m},w}^{S*}$), and $W_{S_{m},p(m)+1}^{S*}=\begin{pmatrix}
\mathbf{0}_{T_{S_{m},p(m)+1}^{S}\times 1}~~(W_{S_{m},p(m)+1}^{S})^{'}
\end{pmatrix}^{'}$ of size $(|V_{\mathcal{G}}|+1)\times T_{S_{m},p(m)+1}^{S}$ denotes independent trees. For homogeneously structured trees, we have $\sum_{m=1}^{q}\sum_{w=1}^{p(m)+1}T_{S_{m},w}^{S}=T^{S}$ and $|V_{W^{S}_{m}}|=|V_{W_{S_{m},w}^{S*}}|$, $w=1,2,\cdots,p(m)+1$, where $|V_{W^{S}_{m}}|$ and $|V_{W_{S_{m},w}^{S*}}|$ denote the number of nodes in matrices $W^{S}_{m}$ and $W_{S_{m},w}^{S*}$, respectively.

\noindent \textbf{Case 1:} Independent trees.

\noindent \textbf{Proposition 5:} Given matrix $W_{S_{m},p(m)+1}^{S*}$ as independent trees, we have $rank(W_{S_{m},p(m)+1}^{S*})=min(|V_{W^{S}_{m}}|,T_{S_{m},p(m)+1}^{S})$, where $|V_{W^{S}_{m}}|$ denotes the number of nodes in matrix $W_{S_{m},p(m)+1}^{S*}$.

\noindent \emph{Proof:} According to the definition of independent matrices, the matrix having the reachability vectors of independent trees from the controller $I^{o}$, i.e. $[R_{TT_{S1}}^{S},R_{TT_{S2}}^{S},\cdots]$, is a structured matrix. For such a structured matrix, we can always find a square sub-matrix of size $min(|V_{W^{S}_{m}}|,T_{S_{m},p(m)+1}^{S})\times min(|V_{W^{S}_{m}}|,T_{S_{m},p(m)+1}^{S})$, whose elements are all non-zero. Therefore, it's obvious that $rank(W_{S_{m},p(m)+1}^{S*})=min(|V_{W^{S}_{m}}|,T_{S_{m},p(m)+1}^{S})$. $\blacksquare$

An illustrative example is given with Fig. 4 (a). The corresponding matrix $W^{S*}_{S_{m},p(m)+1}$ is written as: $W^{S*}_{S_{m},p(m)+1}=\begin{pmatrix}
0 & 1 & a_{1} & a_{1}b_{1} & a_{1}c_{1} & a_{1}d_{1}\\
0 & 1 & a_{2} & a_{2}b_{2} & a_{2}c_{2} & a_{2}d_{2}
\end{pmatrix}^{'}$, whose rank is 2, i.e., $rank(W^{S*}_{S_{m},p(m)+1})=T_{S_{m},p(m)+1}^{S}=2<|V_{W^{S}_{m}}|=5$. More generally, if $T_{S_{m},p(m)+1}^{S}=n>|V_{W^{S}_{m}}|=5$, matrix $W^{S*}_{S_{m},p(m)+1}$ is written as: $W^{S*}_{S_{m},p(m)+1}=\begin{pmatrix}
0 & 0 & \cdots &  0 & 0\\
1 & 1 & \cdots &  1 & 1\\
a_{1} & a_{2} & \cdots &  a_{n} & a_{n}\\
a_{1}b_{1} & a_{2}b_{2}  & \cdots &  a_{n-1}b_{n-1} & a_{n}b_{n}\\
a_{1}c_{1} & a_{2}c_{2}  & \cdots &  a_{n-1}c_{n-1} & a_{n}c_{n}\\
a_{1}d_{1} & a_{2}d_{2}  & \cdots &  a_{n-1}d_{n-1} & a_{n}d_{n}\\
\end{pmatrix}$ and $rank(W^{S*}_{S_{m},p(m)+1})=|V_{W^{S}_{m}}|=5<T_{S_{m},p(m)+1}^{S}=n$.

\noindent \textbf{Case 2:} Interdependent trees.

\noindent \textbf{Proposition 6:} Given matrix $W^{S*}_{S_{m},w}$ as a collection of interdependent trees, we have $rank(W^{S*}_{S_{m},w})=min(|V_{W^{S}_{m}}|-|I_{m,w}|,T_{S_{m},w}^{S}),(w=1,2,\cdots,p(m))$, where $|V_{W^{S}_{m}}|$ denotes the number of nodes, and $|I_{m,w}|$ is the number of same interactions in $W^{S*}_{S_{m},w}$.

\noindent \emph{Proof:} Without loss of generality, we firstly prove the case of two trees as shown in Fig. 4 (b). Here $|I_{m,w}|=3$, i.e. interaction $(B,C,5), (B,D,5)$ and $(B,F,5)$. The corresponding matrix $W^{S*}_{S_{m},w}=\begin{pmatrix}
0 & 1 & a_{1} & a_{1}b & a_{1}c & a_{1}d\\
0 & 1 & a_{2} & a_{2}b & a_{2}c & a_{2}d
\end{pmatrix}^{'}=\begin{pmatrix}
0 & 1 & a_{1} & a_{1}b & a_{1}c & a_{1}d\\
0 & -a_{2}/a_{1} & 0 & 0 & 0 & 0
\end{pmatrix}^{'}$, and it's obvious that the dependence of elements in matrix is caused by the interdependent of trees in some interactions. Thus $rank(W^{S*}_{S_{m},w})=2=|V_{W^{S}_{m}}|-|I_{m,w}|=T_{S_{m},w}^{S}$. More generally, when extending to the case of $n$ trees, $W^{S*}_{S_{m},w}=\begin{pmatrix}
0 & 0 & \cdots & 0 & 0\\
1 & 1 & \cdots & 1 & 1\\
a_{1} & a_{2} & \cdots & a_{n-1} & a_{n}\\
a_{1}b & a_{2}b & \cdots & a_{n-1}b & a_{n}b\\
a_{1}c & a_{2}c & \cdots & a_{n-1}c & a_{n}c\\
a_{1}d & a_{2}d & \cdots & a_{n-1}d & a_{n}d\\
\end{pmatrix}=\begin{pmatrix}
0 & 0 & \cdots & 0 & 0\\
1 & -a_{2}/a_{1} & \cdots & -a_{n-1}/a_{1} & -a_{n}/a_{1}\\
a_{1} & 0 & \cdots & 0 & 0\\
a_{1}b & 0 & \cdots & 0 & 0\\
a_{1}c & 0 & \cdots & 0 & 0\\
a_{1}d & 0 & \cdots & 0 & 0\\
\end{pmatrix}$, and $W^{S*}_{S_{m},n}=2=|V_{W^{S}_{m}}|-|I_{m,n}|<T_{S_{m},n}^{S}=n$. Similarly, for the trees in Fig. 4 (c), $W^{S*}_{S_{m},w}=\begin{pmatrix}
0 & 1 & a_{1} & a_{1}b & a_{1}c & a_{1}d_{1}\\
0 & 1 & a_{1} & a_{1}b & a_{1}c & a_{1}d_{2}\\
\end{pmatrix}^{'}=\begin{pmatrix}
0 & 1 & a_{1} & a_{1}b & a_{1}c & a_{1}d_{1}\\
0 & -a_{2}/a_{1} & 0 & 0 & 0 & (d_{2}-d_{1})a_{2}\\
\end{pmatrix}^{'}$, and $rank(W^{S*}_{S_{m},w})=2=T_{S_{m},w}^{S}<|V_{W^{S}_{m}}|-|I_{m,w}|=3$. Similarly, when extending to the case of $n$ trees, $W^{S*}_{S_{m},w}=\begin{pmatrix}
0 & 0 & \cdots & 0 & 0\\
1 & 1 & \cdots & 1 & 1\\
a_{1} & a_{2} & \cdots & a_{n-1} & a_{n}\\
a_{1}b & a_{2}b & \cdots & a_{n-1}b & a_{n}b\\
a_{1}c & a_{2}c & \cdots & a_{n-1}c & a_{n}c\\
a_{1}d_{1} & a_{2}d_{2} & \cdots & a_{n-1}d_{n-1} & a_{n}d_{n}\\
\end{pmatrix}=\begin{pmatrix}
0 & 0 & \cdots & 0 & 0\\
1 & -a_{2}/a_{1} & \cdots & -a_{n-1}/a_{1} & -a_{n}/a_{1}\\
a_{1} & 0 & \cdots & 0 & 0\\
a_{1}b & 0 & \cdots & 0 & 0\\
a_{1}c & 0 & \cdots & 0 & 0\\
a_{1}d_{1} & (d_{2}-d_{1})a_{2} & \cdots & (d_{n-1}-d_{1})a_{n-1} & (d_{n}-d_{1})a_{n}\\
\end{pmatrix}$, and $W^{S*}_{S_{m},w}=3=|V_{W^{S}_{m}}|-|I_{m,w}|<T_{S_{m},w}^{S}=n$. $\blacksquare$

\noindent \textbf{Theorem 2:} Given matrices $W^{S*}_{S_{m},w}$, $w=1,2,\cdots,p(m)+1$, as homogeneously structured trees, we have
\begin{equation}
rank(W^{S}_{m})=min\{min[\sum_{w=1}^{p(m)}(rank(W^{S*}_{S_{m},w})),max_{w=1}^{p(m)}\{|V_{W^{S}_{m}}|-|I_{m,w}|\}]
+rank(W^{S*}_{S_{m},p(m)+1}),|V_{W^{S}_{m}}|\}
\end{equation}
where $|V_{W^{S}_{m}}|$ is the number of nodes, and $|I_{m,w}|$ is the number of same interactions in $W^{S*}_{S_{m},w}$, $w=1,2,\cdots,p(m)$.

\noindent \emph{Proof:} The outsider function $min\{~\}$ ensures that the rank of matrix $W^{S}_{m}$ never exceeds the number of independent rows, i.e., the number of nodes in matrix $W^{S}_{m}$. Next we focus on the number of independent columns. From the proof of \emph{Proposition 5}, we know there always exists a structured square matrix of size $min(|V_{W^{S}_{m}}|,T_{S_{m},p(m)+1})\times min(|V_{W^{S}_{m}}|,T_{S_{m},p(m)+1})$ in matrix $W^{S*}_{S_{m},p(m)+1}$, so there always exists $min(|V_{W^{S}_{m}}|,T_{S_{m},p(m)+1})$ independent columns compared with interdependent matrix $W^{id}=\begin{pmatrix}
\mathbf{0}_{1\times T_{S_{m},1}^{S}} & \cdots & \mathbf{0}_{1\times T_{S_{m},p(m)}^{S}}\\
W_{S_{m},1}^{S} & \cdots & W_{S_{m},p(m)}^{S}
\end{pmatrix}$, which means matrix $W_{S_{m},p(m)+1}^{S*}$ always contributes $rank(W_{S_{m},p(m)+1}^{S*})$ to matrix $W^{S}_{m}$, i.e., $rank(W_{S_{m},p(m)+1}^{S*})$ in Eq. (14). Now we focus on the part of $min[\sum_{w=1}^{p(m)}(rank(W^{S*}_{S_{m},w})),max_{w=1}^{p(m)}\{|V_{W^{S}_{m}}|-|I_{m,w}|\}]$, which deals with the rank of all interdependent trees, i.e. the rank of matrix $W^{id}$. Without loss of generality, for trees shown in Fig. 4 (b) and (c), we have $\begin{pmatrix}W^{S}_{S_{m},1}~W^{S}_{S_{m},2}\end{pmatrix}=\begin{pmatrix}
0 & 0 & \vline & 0 & 0 \\
1 & 1 & \vline & 1 & 1\\
a_{1} & a_{2} & \vline & a_{1} & a_{2}\\
a_{1}b & a_{2}b & \vline & a_{1}b & a_{2}b\\
a_{1}c & a_{2}c & \vline & a_{1}c & a_{2}c\\
a_{1}d & a_{2}d & \vline & a_{1}d_{1} & a_{2}d_{2}
\end{pmatrix}=\begin{pmatrix}
0 & 0 & \vline & 0 & 0 \\
1 & -a_{2}/a_{1} & \vline & 0 & -a_{2}/a_{1}\\
a_{1} & 0 & \vline & 0 & 0\\
a_{1}b & 0 & \vline & 0 & 0\\
a_{1}c & 0 & \vline & 0 & 0\\
a_{1}d & 0 & \vline & a_{1}(d_{1}-d) & a_{2}(d_{2}-d)
\end{pmatrix}$, and $rank(W^{S}_{S_{m},1}~W^{S}_{S_{m},2})=3<rank(W^{S}_{S_{m},1})+rank(W^{S}_{S_{m},2})=4$. More generally, when extending to the case of $n$ trees, we similarly have $\begin{pmatrix}W^{S}_{S_{m},1}~~W^{S}_{S_{m},2}\end{pmatrix}=
\begin{pmatrix}
\begin{array}{ccccccccccc}
0 & 0 & \cdots & 0 & 0 & \vline & 0 & 0 & \cdots & 0 & 0\\
1 & 1 & \cdots & 1 & 1 & \vline & 1 & 1 & \cdots & 1 & 1\\
a_{1} & a_{2} & \cdots & a_{n-1} & a_{n} & \vline & a_{1} & a_{2} & \cdots & a_{n-1} & a_{n}\\
a_{1}b & a_{2}b & \cdots & a_{n-1}b & a_{n}b & \vline & a_{1}b & a_{2}b & \cdots & a_{n-1}b & a_{n}\\
a_{1}c & a_{2}c & \cdots & a_{n-1}c & a_{n}c & \vline & a_{1}c_{1} & a_{2}c_{2} & \cdots & a_{n-1}c_{n-1} & a_{n}c_{n}\\
a_{1}d & a_{2}d & \cdots & a_{n-1}d & a_{n}d & \vline & a_{1}d_{1} & a_{2}d_{2} & \cdots & a_{n-1}d_{n-1} & a_{n}d_{n}
\end{array}
\end{pmatrix}=
\begin{pmatrix}
\begin{array}{ccccccccccc}
0 & 0 & \cdots & 0 & 0 & \vline & 0 & 0 & \cdots & 0 & 0\\
1 & -a_{2}/a_{1} & \cdots & -a_{n-1}/a_{1} & -a_{n}/a_{1} & \vline & 0 & -a_{2}/a_{1} & \cdots & -a_{n-1}/a_{1} & -a_{n}/a_{1}\\
a_{1} & 0 & \cdots & 0 & 0 & \vline & 0 & 0 & \cdots & 0 & 0\\
a_{1}b & 0 & \cdots & 0 & 0 & \vline & 0 & 0 & \cdots & 0 & 0\\
a_{1}c & 0 & \cdots & 0 & 0 & \vline & a_{1}(c_{1}-c) & a_{2}(c_{2}-c) & \cdots & a_{n-1}(c_{n-1}-c) & a_{n}(c_{n}-c)\\
a_{1}d & 0 & \cdots & 0 & 0 & \vline & a_{1}(d_{1}-d) & a_{2}(d_{2}-d) & \cdots & a_{n-1}(d_{n-1}-d) & a_{n}(d_{n}-d)
\end{array}
\end{pmatrix}$, and $rank(W^{S}_{S_{m},1}~W^{S}_{S_{m},2})=4<rank(W^{S}_{S_{m},1})+rank(W^{S}_{S_{m},2})=5$. When $\sum_{w=1}^{p(m)}(rank(W^{S*}_{S_{m},w}))\leq max_{w=1}^{p(m)}(|V_{W^{S}_{m}}|-|I_{m,w}|)$, it's easy to verify that $rank(W^{S}_{m})=\sum_{w=1}^{p(m)}(rank(W^{S*}_{S_{m},w}))$. So the rank of matrix $W^{id}$ is $min[\sum_{w=1}^{p(m)}(rank(W^{S*}_{S_{m},w})),max_{w=1}^{p(m)}\{|V_{W^{S}_{m}}|-|I_{m,w}|\}]$. $\blacksquare$

With Eq. (12) and \emph{Theorem 2}, we directly give the following \emph{Lemma 1} for homogeneously structured trees.

\noindent \textbf{Lemma 1:} Given matrices $W^{S}_{m}$, $m=1,2,\cdots,q$, as collections of homogeneously structured trees and $S^{S}_{M(o)}$ as the maximum-structurally controllable subspace of homogeneously structured trees, we have
\begin{equation}
max_{m=1}^{q}rank(W^{S}_{m})\leq S^{S}_{M(o)}=rank(W^{S})\leq \sum_{m=1}^{q}rank(W^{S}_{m})
\end{equation}


With \emph{Theorem 1}, \emph{Theorem 2} and \emph{Lemma 1} above, we straightly get \emph{Theorem 3}:

\noindent \textbf{Theorem 3:} Given $S^{D}_{M(o)}$ and $S^{S}_{M(o)}$ as the maximum controlled subspace of heterogeneously structured and homogeneously structured temporal trees in Eq. (11) and (15), respectively, we have
\begin{equation}
max(S^{D}_{M(o)},S^{S}_{M(o)})\leq S_{M(o)}\leq S^{D}_{M(o)}+S^{S}_{M(o)}
\end{equation}\\
\noindent \textbf{Numerical Simulations} We firstly verify the feasibility and reliability of \emph{Theorem 3}. As shown in Fig. 5, four different networks with
sizes of 40, 60, 80 and 100 are studied, respectively. For each of the four networks, we randomly generate an interaction between a pair of nodes with probability 0.002, and repeat it for all the $N(N-1)/2$ pairs of nodes at a specified time point. Repeat this process for 100 rounds at 100 different time points , i.e. $t=1,2,\cdots,100$. As shown in Fig. 5, all the calculated values of controlling centrality of the four networks (denoted as 'Calculated') are between the upper and lower bounds (denoted as 'Upper Bound' and 'Lower Bound') given by our analytical results in Eq. (16). Besides, the gaps (numerical calculations) between upper and lower bounds are very minor in these artificial networks.


We further investigate three empirical datasets, i.e., 'HT09', 'SG-Infectious' and 'Fudan WIFI' (Details of the datasets see Methods)~\cite{isella2011jtb,zhang2012,zhang2012epl,zhang2013chaos}. For the dataset of 'HT09', two temporal networks are generated: i) a temporal network (113 nodes and 9865 interactions) with all nodes and interactions within record of dataset, denoted as 'all range', ii) a temporal network (73 nodes and 3679 interactions) with nodes and interactions after removing the most powerful nodes (nodes with the largest controlling centrality) in the temporal network of i), denoted as 'removed'. For the dataset of 'SG-Infectious', three temporal networks are generated: i) a temporal network (1321 nodes and 20343 interactions) with nodes and interactions recorded in the first week, denoted as 'Week 1', ii) a temporal network (868 nodes and 13401 interactions) with nodes and interactions recorded in the second week, denoted as 'Week 2', iii) a temporal network (2189 nodes and 33744 interactions) with nodes and interactions recorded in the first two weeks, denoted as 'Week 1\&2'. For the dataset of 'Fudan WIFI', three temporal networks are generated: i) a temporal network (1120 nodes and 12833 interactions) with nodes and interactions recorded in the first day, denoted as 'Day 1', ii) a temporal network (2250 nodes and 25772 interactions) with nodes and interactions recorded in the second day, denoted as 'Day 2', iii) a temporal network (1838 nodes and 27810 interactions) with nodes and interactions recorded at Access Point No.713, denoted as '713 point'. With these three types of eight temporal networks, we calculate their upper and lower bounds of controlling centrality given by our analytical results. The aggregated degree of a node in Figs. 6 and 7 is the number of neighbored nodes whom it interacts within the corresponding temporal network. As shown in Fig. 6, although the sizes of these networks range from 73 to 2250, the gaps of the upper and lower bounds remain very tiny, indicating the feasibility and reliability of Eq. (16) in both artificial (refer to Fig. 5) and empirical networks. Further more, Fig. 7 shows us the positive relationship between the aggregated degree and controlling centrality of nodes. When removing the most powerful nodes (nodes with the largest controlling centrality), as shown in Fig. 7 (a), and considering temporal networks with different time scales and types, as shown in Fig. 7 (b) and (c), the observed positive relationship remains unchanged. This indicates the robustness of this relationship of temporal network, regardless of the structural destructions or time evolutions of the network.

Besides, Fig. 8 focuses on the datasets of 'SG-Infectious' and 'Fudan WIFI' to visualize the distribution of controlling centrality of different temporal networks. The scale-free distribution of node's controlling centrality is virtually independent of the time period and network scale, which is similar to the distribution of node's activity potential~\cite{perra2012sr}. However, these two studied datasets are inherently different. The dataset of 'SG-Infectious' collected the attendee's temporal activity information during an exhibition, where the attendee generally do not appear again after the visit. Therefore, the interactions among nodes in the temporal networks generated  from 'SG-Infectious' present more randomness than those of 'Fudan WIFI', while the latter presents weekly rhythm of the scheduled campus activities in a university.

\noindent \textbf{\large Methods}

\noindent \textbf{Notation}
The symbols used in the main text are summarized in Table II.
\begin{table}
\begin{tabular}{|c|c|c|c|}
  \hline
  Notations & Description \\
  \hline
  $G^{t}$ & the $t^{th}$ formation of temporal network $\mathcal{G}(V_{\mathcal{G}},E_{\mathcal{G}})$ \\
  \hline
  $V$ and $|V|$ & the set of nodes and the cardinality of set $V$\\
  \hline
  $A_{t}$ & the adjacency matrix of graph $G^{t}$ \\
  \hline
  $(A)^{'}$ & the transpose of adjacency matrix $A$\\
  \hline
  $(A)^{k}$ & the $k^{th}$ power of adjacency matrix $A$ \\
  \hline
  $\{A\}_{i,j}$ & an element of matrix $A$ with position $i$ (row index) and $j$ (column index)\\
  \hline
  $\{A\}_{i,\forall}$ & the $i^{th}$ row of matrix $A$\\
  \hline
  $I^{o}$ & the controller located on node $o$ of temporal network $\mathcal{G}(V_{\mathcal{G}},E_{\mathcal{G}})$\\
  \hline
  $Q_{t}$ & dynamic communicability matrix of temporal network $\mathcal{G}(V_{\mathcal{G}},E_{\mathcal{G}})$ at time $t$\\
  \hline
  $W^{*}$ & reachability matrix of input signal within the temporal network $\mathcal{G}(V_{\mathcal{G}},E_{\mathcal{G}})$\\
  \hline
  $R_{TT}$ & reachability vector of input signal within a temporal tree $TT$\\
  \hline
  $R_{TT}^{D}$ & reachability vector of input signal within heterogeneously structured temporal tree $TT$\\
  \hline
  $R_{TT}^{S}$ & reachability vector of input signal within homogeneously structured temporal tree $TT$\\
  \hline
  $W^{R}$ & reachability matrix of input signal within temporal trees extracted from\\
  ~~ &  temporal network $\mathcal{G}(V_{\mathcal{G}},E_{\mathcal{G}})$\\
  \hline
  $W^{D}$ & reachability matrix of input signal within heterogeneously structured temporal trees\\
  \hline
  $W^{S}$ & reachability matrix of input signal within homogeneously structured temporal trees\\
  \hline
  $S_{M(o)}$ & the maximum controlled subspace of temporal network $\mathcal{G}(V_{\mathcal{G}},E_{\mathcal{G}})$\\
  ~~ & with single controller located on $o$\\
  \hline
  $S_{M(o)}^{D}$ & the maximum controlled subspace of heterogeneously structured temporal trees\\ 
  ~~ & with single controller located on $o$\\
  \hline
  $S_{M(o)}^{S}$ & the maximum controlled subspace of homogeneously structured temporal trees\\
  ~~ & with single controller located on $o$\\
  \hline
\end{tabular}
\caption{Notations in the paper}
\end{table}

\noindent \textbf{Controlling Centrality} With a sampling interval properly chosen, we write Eq. (3) as follow:
\begin{equation}
\frac{x(k+1)-x(k)}{T_{k+1}}=A_{k+1}^{'}x(k)+b^{(o)}u(k)
\end{equation}
Generally, $T_{k+1}\neq T_{k}$, where $T_{k+1}=t_{k+1}-t_{k}$ is the sampling interval.
From Eq. (17), we get the recursive relationship of two neighboring states as:
\begin{equation}
x(k+1)=G_{k+1}x(k)+H_{k+1}u(k)
\end{equation}
Where $G_{k+1}=I+T_{k+1}A_{k+1}^{'}$, $H_{k+1}=T_{k+1}b^{(o)}$, $A_{k+1}^{'}$ and $I$ are the transpose of the adjacency matrix of the $(k+1)$th graph and the identity matrix, respectively.
Define
\begin{equation}
W_{c}=[G_{T}\cdots G_{2}H_{1}, \cdots, G_{T}H_{T-1}, H_{T}]
\end{equation}
and the final state is written as:
\begin{equation}
\begin{split}
x(T)& =[x_{1}(T),\cdots,x_{N}(T)]^{'}\\
& =[G_{T}\cdots G_{1}]\cdot[x_{1}(0),\cdots,x_{N}(0)]^{'}\\
& +W_{c}\cdot[u(0),u(1),\cdots,u(T-1)]^{'}
\end{split}
\end{equation}
If there exists a sequence of inputs denoted as $[u(0),u(1),\cdots,u(T-1)]^{'}$ such that $[x_{1}(T),\cdots,x_{N}(T)]^{'}=[0,\cdots,0]^{'}$ in Eq. (20), then the temporal network is structurally controllable at time point $t_{0}$, i.e. $rank(W_{c})=N$. Otherwise, we may split $x(T)$ into two parts, written as:
\begin{equation}
\begin{array}{cccccccc}
\begin{split}
x(T)& =[x_{1}(T),\cdots,x_{k}(T),x_{k+1}(T),\cdots,x_{N}(T)]^{'}\\
& =[G_{T}\cdots G_{1}]\cdot[x_{1}(0),\cdots,x_{k}(0),x_{k+1}(0),\cdots,x_{N}(0)]^{'}\\
& +W_{c}\cdot[u(0),u(1),\cdots,u(T-1)]^{'}
\end{split}
\end{array}
\end{equation}
and if there exists a sequence of inputs denoted as $[u(0),u(1),\cdots,u(T-1)]^{'}$ such that $[x_{1}(T),\cdots~x_{k}(T)]^{'}=[0~\cdots~0]^{'}$ in Eq. (21), then the $k$ subspace of the network is structurally controllable at time point $t_{0}$, which is equivalent to the condition $rank(W_{c})=k$. Therefore, we define controlling centrality as
\begin{equation}
S_{M(o)}=rank(W_{c})
\end{equation}
i.e. the maximum dimension of controllable subspace, as a measure of node $o$'s ability to structurally control the network: if $S_{M(o)}=N$, then node $o$ alone can structurally control the whole network. Any value of $S_{M(o)}$ less than $N$ provides the maximum dimension of the subspace $o$ can structurally control.\\

\noindent \textbf{Datasets} We mainly investigate three temporal networks with three empirical data sets in this paper. The first data was collected during the ACM Hypertext 2009 conference, where the 'SocioPatterns' project deployed the Live Social Semantics applications. The conference attendees volunteered to wear radio badges which monitored their face-to-face interactions and we name this data as 'HT09'. The second is a random data set containing the daily dynamic contacts collected during the art-science exhibition 'INFECTIOUS: STAY AWAY' which took place at the Science Gallery in Dublin, Ireland, and we name it as 'SG-Infectious'. These two data are both available from the website of 'SocioPatterns'~\cite{isella2011jtb} (http://www.sociopatterns.org). The third data set was collected from Fudan University during the 2009-2010 fall semester (3 whole months), which is named as 'FudanWIFI'~\cite{zhang2012,zhang2012epl,zhang2013chaos}. In this data set, each student/teacher/visiting scholar has a unique account to access the Campus WiFi system, which automatically records the device' MAC addresse, the MAC address of the accessed WiFi access point (APs), and the connecting (disconnecting) time as well. Table III summaries some characteristics of the aforementioned three empirical datasets.\\
\begin{table}
\begin{tabular}{|c|c|c|c|}
  \hline
  ~~~~~ & HT09 & SG-Infectious & FudanWIFI \\
  \hline
  Area & Conference & Mesume & Campus \\
  \hline
  Technology & RFID & RFID & WiFi \\
  \hline
  Collection duration & 3 days & 62 days & 84 days \\
  \hline
  Number of individuals & 113 & 10970 & 17897\\
  \hline
  Number of contacts & 9865 & 198198 & 884800\\
  \hline
  Spatial resolution(meters) & $<$2 & $<$2 & $<$8\\
  \hline
  Types of contacts & Strangers with repeat & Acquaintances without repeat & Acquaintances with repeat\\
  \hline
\end{tabular}
\caption{Characteristics of the three empirical datasets}
\end{table}

\clearpage

\noindent \textbf{Acknowledgments} \\
This work was partly supported by National Key Basic Research and Development Program (No. 2010CB731403), the National Natural Science Foundation (No. 61273223), the Research Fund for the Doctoral Program of Higher Education (No. 20120071110029) and the Key Project of National Social Science Fund (No. 12\&ZD18) of China.

\noindent \\ \textbf{Author contributions} \\
YJP \& XL planned the study; YJP performed the experiments; YJP \& XL analyzed the data and wrote the manuscript.

\noindent \\ \textbf{Additional Information} \\
{\em Supplementary Information} for " ".

\noindent \\ \textbf{Competing financial interests} \\
The authors declare no competing financial interests.

\vspace{2cm}

\noindent  \textbf{\Large Figure Legends} \\

\clearpage

\begin{figure}
\begin{center}
\includegraphics[width=5in]{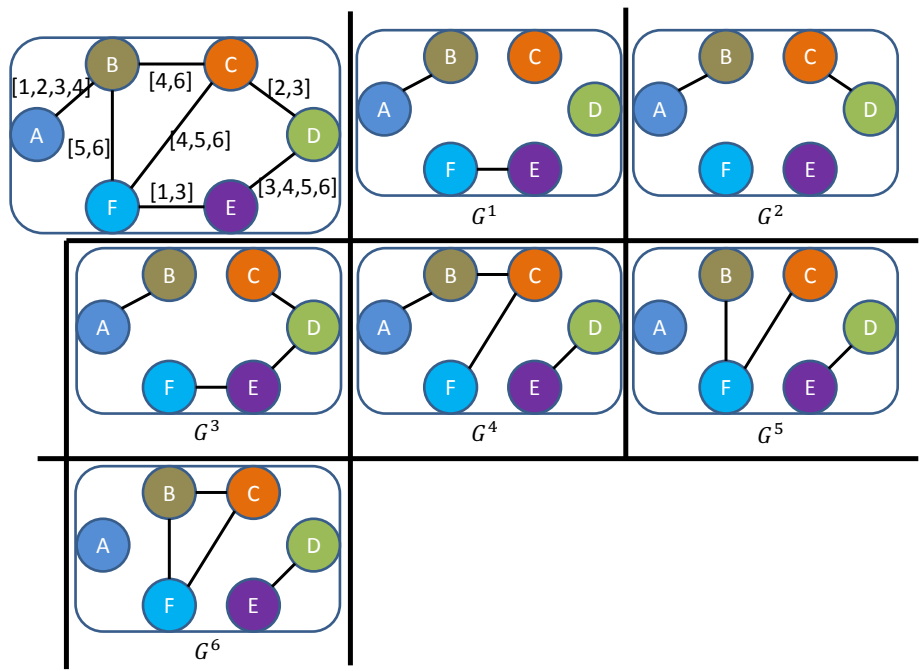}
\end{center}
\caption{\label{fig.1}\textbf{The sequence of graphs representation of the contacts in Table I.} In each discrete time point, the network has a different formation shown as $G^{1}, \cdots, G^{6}$.}
\end{figure}

\begin{figure}
\begin{center}
\includegraphics[width=5in]{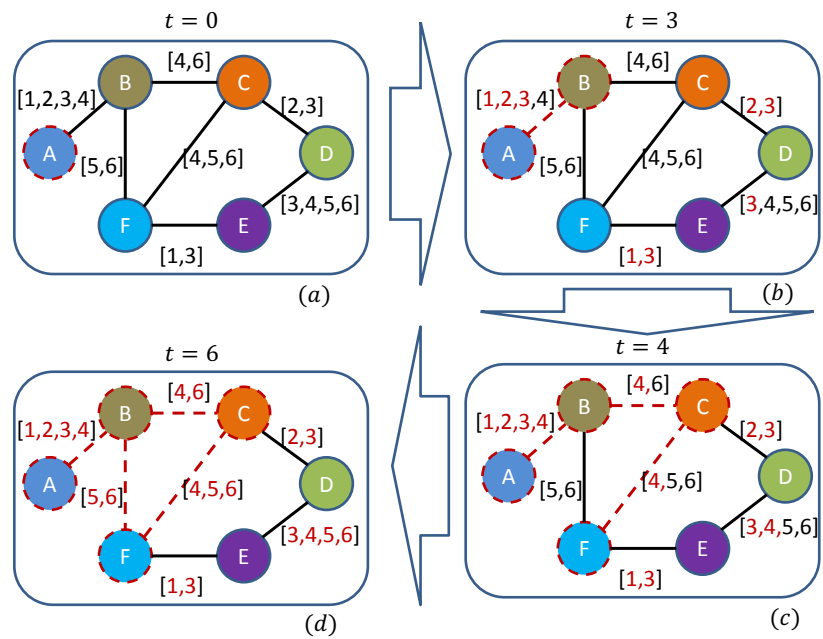}
\end{center}
\caption{\label{fig.1}\textbf{The illustration of information propagation on a temporal network.} (a), (b), (c) and (d) denote different networks at different time points, respectively. Red (gray) time points on edges denote the elapsed time, and the black (dark) time points denote the forthcoming time.}
\end{figure}

\begin{figure}[!ht]
\begin{center}
\includegraphics[width=15cm]{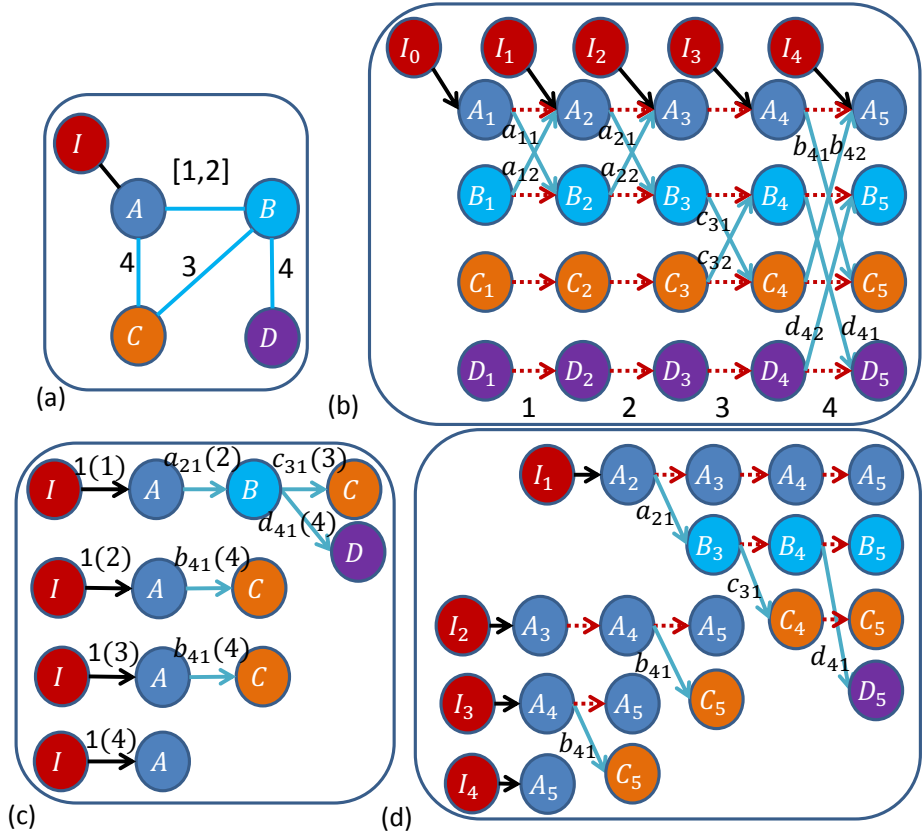}
\end{center}
\caption{\label{fig.2}\textbf{The illustration of transformation of a temporal network to a static one.} (a) Temporal Network with a single controller located on node \emph{A}, (b) The Time-Ordered Graph (TOG), (c) The temporal trees of (a) at time points 1, 2, 3 and 4, (d) the BFS spanning trees of TOG. The red (dashed), black (dark) and blue (light) lines stand for the flows of time order, the connection with the single controller and the interactions of individuals, respectively. The numbers with parenthesis in (c) denote time stamps. Weights of interactions (the blue ones) are labeled by characters $a_{11},a_{12},\cdots,d_{41},d_{42}$ in (b), (c) and (d), and without loss of generality, we denote the weight of other edges (the red and black ones) as "1".}
\end{figure}

\begin{figure}[!ht]
\begin{center}
\includegraphics[width=5in]{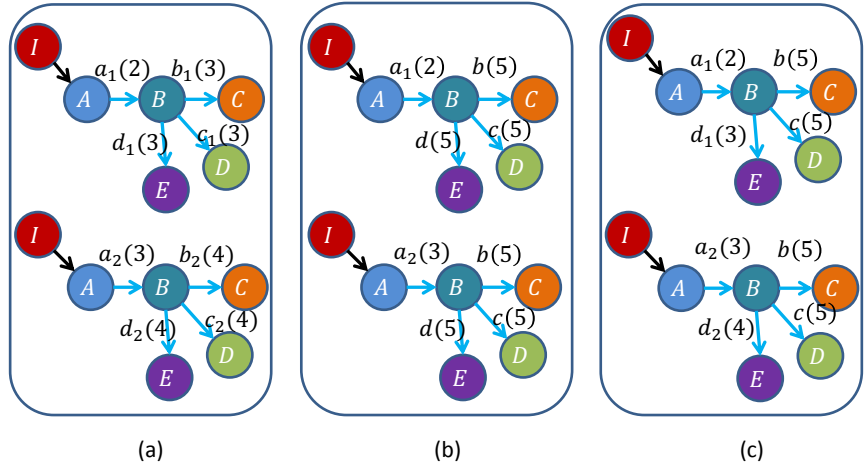}
\end{center}
\caption{\label{fig.7} \textbf{Three examples of the homogeneously structured temporal trees.} (a) Independent trees, (b) and (c) Interdependent trees. For the two homogeneously structured trees in (b), there are three same interactions, i.e (B,C,5), (B,D,5) and (B,E,5), but there are only two such interactions, i.e (B,C,5) and (B,D,5), for the trees in (c). The trees in (b) and (c) are both interdependent according to our definition. The numbers in parenthesis denote active time points of interactions and characters $a_{1}, a_{2}, b, b_{1}, b_{2}, c, c_{1}, c_{2}, d, d_{1}$ and $d_{2}$ denote the weights of interactions.}
\end{figure}

\begin{figure}[!ht]
\begin{center}
\includegraphics[width=16cm]{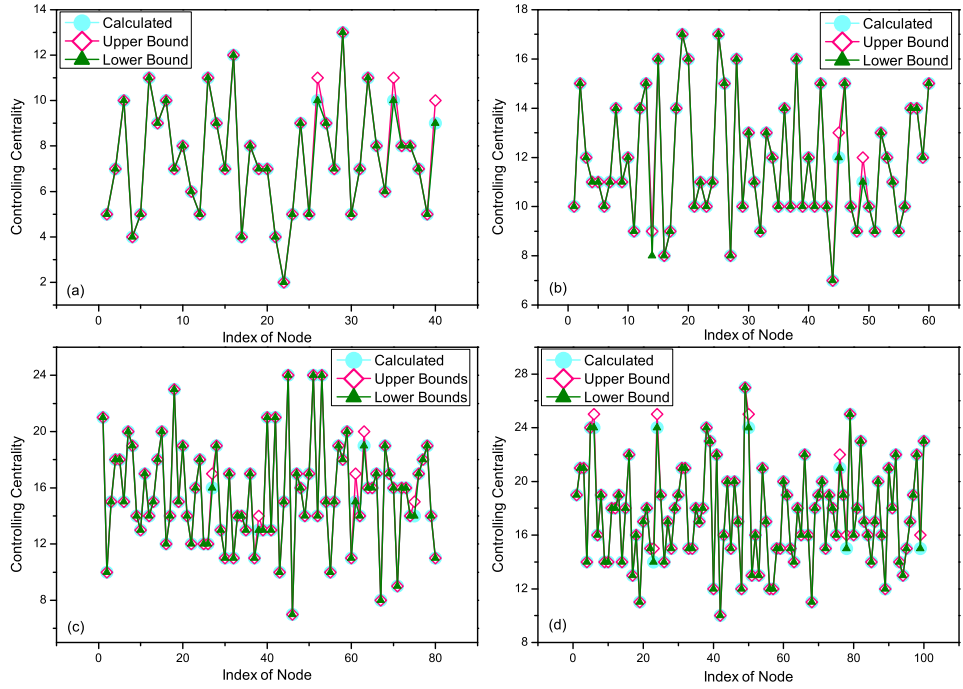}
\end{center}
\caption{\label{fig.4}\textbf{Controlling centrality of artificial networks.} (a), (b), (c) and (d) denote network with 40, 60, 80 and 100 nodes, respectively. For each of the four networks, we randomly generate an interaction between a pair of nodes with probability 0.002, and repeat it for all the $N(N-1)/2$ pairs of nodes at a specified time point. repeat this process for 100 rounds at 100 different time points , i.e. $t=1,2,\cdots,100$. The value of controlling centrality, denoted as 'Calculated', is straightly calculated by the computation of matrix $W_{c}$ in Eq. (19), and the upper and lower bounds, denoted as 'Upper Bound' and 'Lower Bound', respectively, are given by the analytical results in Eq. (16).}
\end{figure}

\begin{figure}[!ht]
\begin{center}
\includegraphics[width=16cm]{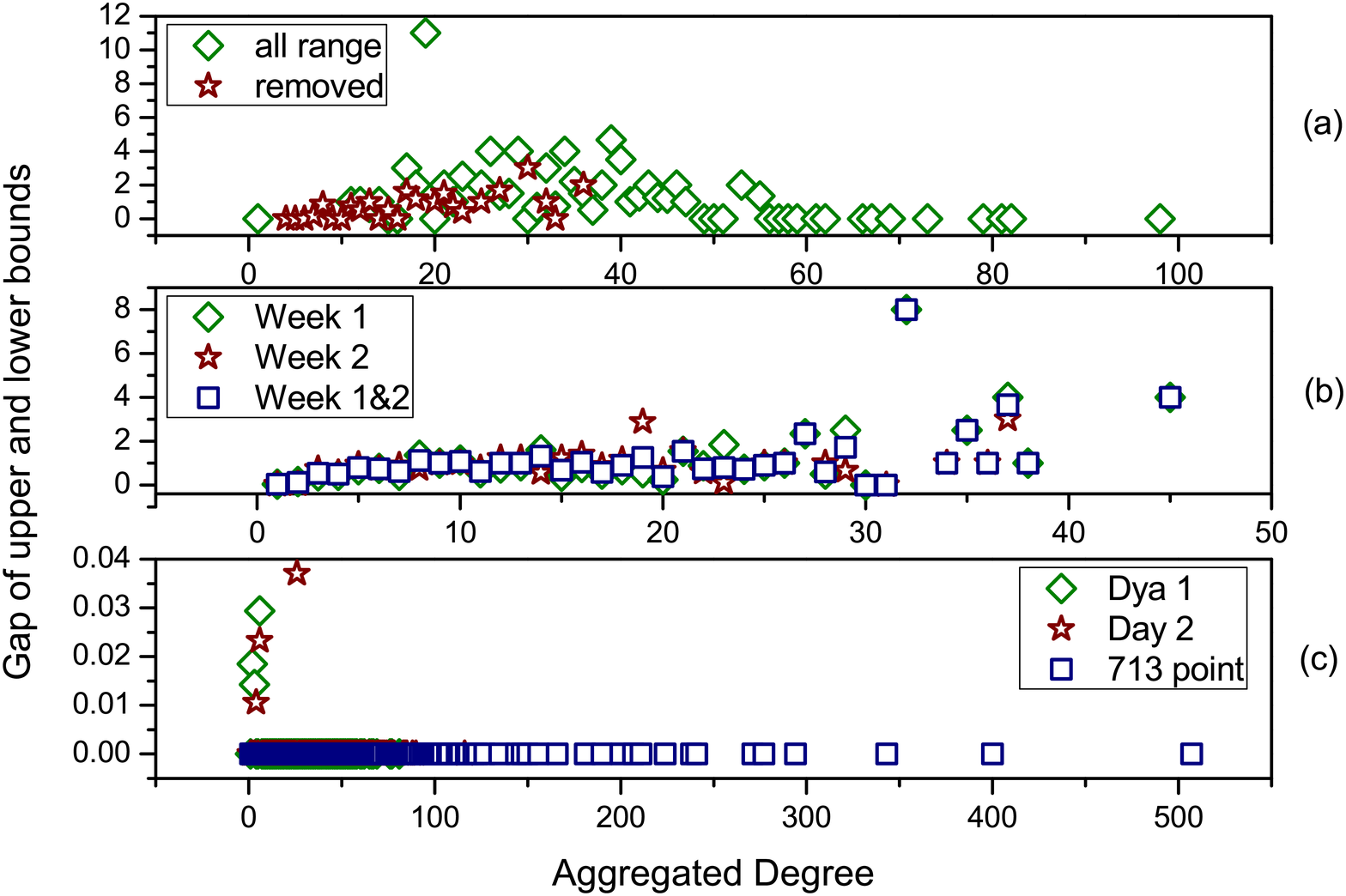}
\end{center}
\caption{\textbf{The gap of upper and lower bounds of controlling centrality.} (a) HT09 (b) SG-Infectious (c) Fudan WIFI. For the dataset of 'HT09', two temporal networks are generated: i) a temporal network (113 nodes and 9865 interactions) with all nodes and interactions within record of dataset, denoted as 'all range', ii) a temporal network (73 nodes and 3679 interactions) with nodes and interactions after removing the most powerful nodes (nodes with the largest controlling centrality) in the temporal network of i), denoted as 'removed'. For the dataset of 'SG-Infectious', three temporal networks are generated: i) a temporal network (1321 nodes and 20343 interactions) with nodes and interactions recorded in the first week, denoted as 'Week 1', ii) a temporal network (868 nodes and 13401 interactions) with nodes and interactions recorded in the second week, denoted as 'Week 2', iii) a temporal network (2189 nodes and 33744 interactions) with nodes and interactions recorded in the first two weeks, denoted as 'Week 1\&2'. For the dataset of 'Fudan WIFI', three temporal networks are generated: i) a temporal network (1120 nodes and 12833 interactions) with nodes and interactions recorded in the first day, denoted as 'Day 1', ii) a temporal network (2250 nodes and 25772 interactions) with nodes and interactions recorded in the second day, denoted as 'Day 2', iii) a temporal network (1838 nodes and 27810 interactions) with nodes and interactions recorded at Access Point No.713, denoted as '713 point'. The upper and lower bounds of the controlling centrality are given by analytical results in the main text, and the gap is given by the absolute value of the difference of the upper and lower bounds. The aggregated degree of a node is the number of neighbored nodes whom it interacts within the corresponding temporal network. All the gaps are minor when compared with the sizes of these temporal networks.}
\end{figure}

\begin{figure}[!ht]
\begin{center}
\includegraphics[width=16cm]{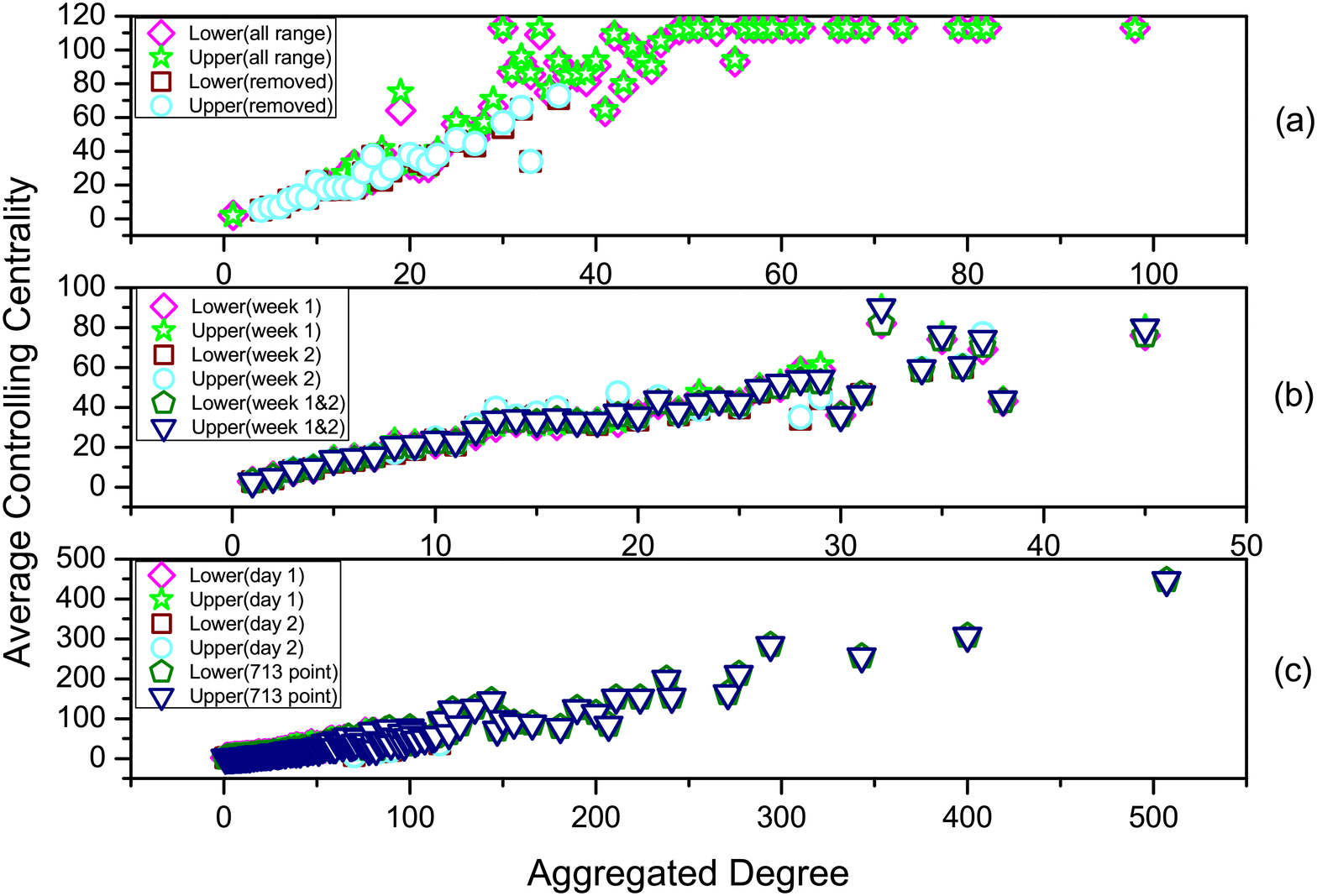}
\end{center}
\caption{\label{fig.5}\textbf{The relationship between node's aggregated degree and the average controlling centrality.} (a) HT09 (b) SG-Infectious (c) FudanWIFI. All the temporal networks are the same as those in Fig. 6. Each point in this figure is an average controlling centrality of nodes with the same aggregated degree, and there's a positive relationship between the aggregated degree and its controlling centrality, even with some structural destructions or time evolutions.}
\end{figure}

\begin{figure}[!ht]
\begin{center}
\includegraphics[width=16cm]{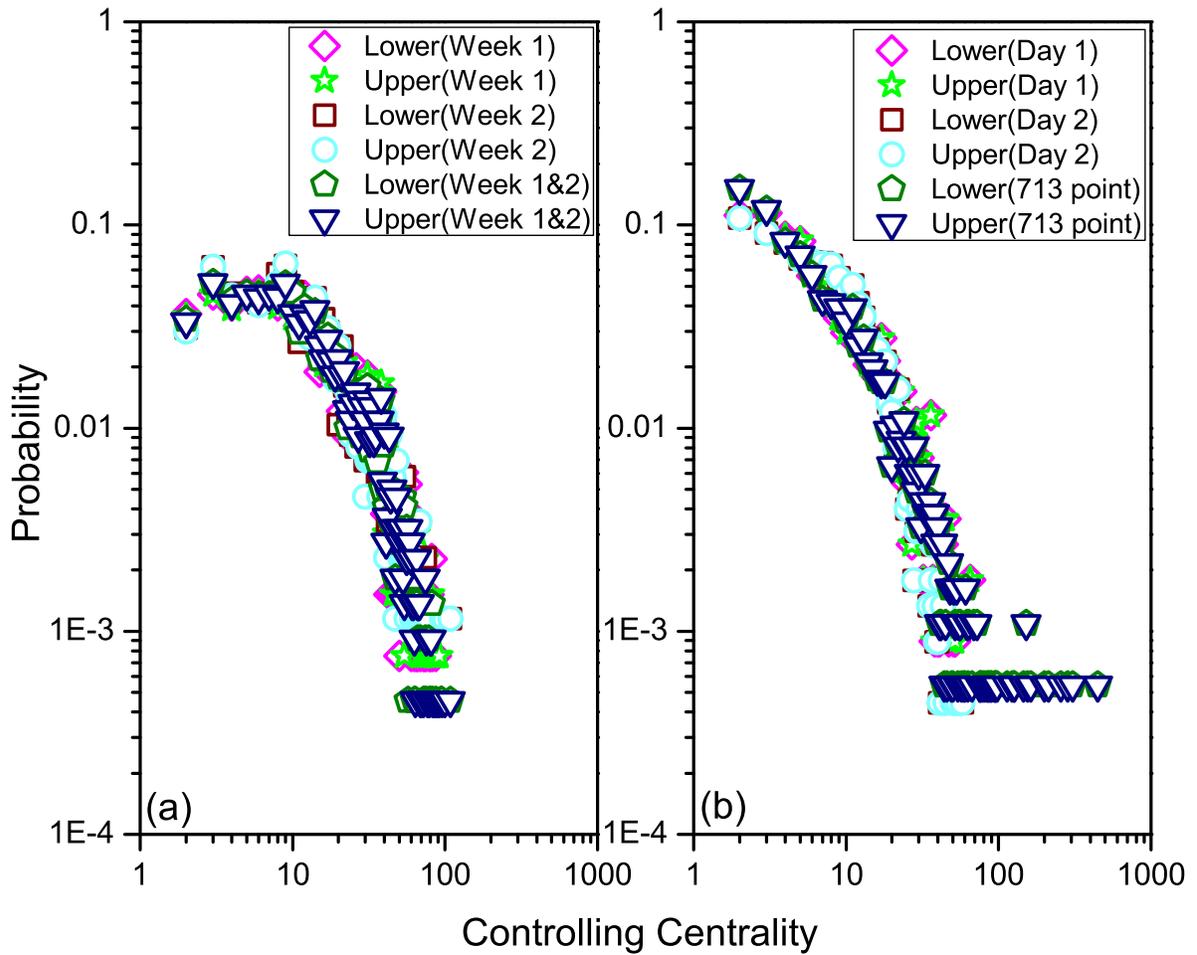}
\end{center}
\caption{\label{fig.6} \textbf{The distribution of node's controlling centrality.} (a) Temporal networks generated by the dataset of 'SG-Infectious' (b) Temporal networks generated by the dataset of 'Fudan WIFI'. For each dataset, three different temporal networks are generated within different time scales, denoted as 'Week 1', 'Week 2' and 'Week 1\&2' for SG-Infectious and 'Day 1', 'Day 2' and '713 point' for Fudan WIFI, respectively.}
\end{figure}


\end{document}